\documentclass[useAMS,usenatbib]{mn2e}
\usepackage{epsfig}
\usepackage{amsmath}

\newcommand{\be}{\begin{equation}}
\newcommand{\beq}{\begin{equation}}
\newcommand{\ba}{\begin{eqnarray}}
\newcommand{\ee}{\end{equation}}
\newcommand{\eeq}{\end{equation}}
\newcommand{\ea}{\end{eqnarray}}

\newcommand{\apj}{ApJ}
\newcommand{\apjl}{ApJL}
\newcommand{\mnras}{MNRAS}
\newcommand{\aj}{AJ}

\def\lsim{~\rlap{$<$}{\lower 1.0ex\hbox{$\sim$}}}

\def\gsim{~\rlap{$>$}{\lower 1.0ex\hbox{$\sim$}}}

\title[Galactic hydrogen and 21-cm fluctuations]{The Impact of HI in Galaxies on 21-cm Intensity Fluctuations During the Reionisation Epoch}

\author[Wyithe, Warszawski, Geil, Oh]{J.S.B. Wyithe$^1$\thanks{swyithe@unimelb.edu.au}, L. Warszawski$^1$, P.M. Geil$^1$, S. Peng Oh$^2$ \\
$^1$School of Physics, University of Melbourne, Parkville, VIC 3010, Australia\\
$^2$Department of Physics, University of California, Santa Barbara, CA 93106, USA.\\}

\begin{document}


\maketitle

\label{firstpage}
\begin{abstract}

  We investigate the impact of neutral hydrogen (H\,{\sevensize I}) in
  galaxies on the statistics of 21-cm fluctuations using analytic and
  semi-numerical modelling.  Following the reionisation of hydrogen
  the H\,{\sevensize I} content of the Universe is dominated by damped
  absorption systems (DLAs), with a cosmic density in H\,{\sevensize
    I} that is observed to be constant at a level equal to $\sim2\%$
  of the cosmic baryon density from $z\sim1$ to $z\sim5$. We show that
  extrapolation of this constant fraction into the reionisation epoch
  results in a reduction of 10-20\% in the amplitude of 21-cm
  fluctuations over a range of spatial scales. The assumption of a
  different percentage during the reionisation era results in a
  proportional change in the 21-cm fluctuation amplitude. We find that
  consideration of H\,{\sevensize I} in galaxies/DLAs reduces the
  prominence of the H\,{\sevensize II} region induced {\em shoulder}
  in the 21-cm power spectrum (PS), and hence modifies the scale
  dependence of 21-cm fluctuations. We also estimate the 21cm-galaxy
  cross PS, and show that the cross PS changes sign on scales
  corresponding to the H\,{\sevensize II} regions. From consideration
  of the sensitivity for forthcoming low-frequency arrays we find that
  the effects of HI in galaxies/DLAs on the statistics of 21-cm
  fluctuations will be significant with respect to the precision of a
  PS or cross PS measurement. In addition, since overdense regions are
  reionised first we demonstrate that the cross-correlation between
  galaxies and 21-cm emission changes sign at the end of the
  reionisation era, providing an alternative avenue to pinpoint the
  end of reionisation. The sum of our analysis indicates that the
  H\,{\sevensize I} content of the galaxies that reionise the universe
  will need to be considered in detailed modelling of the 21-cm
  intensity PS in order to correctly interpret measurements from
  forthcoming low-frequency arrays.  \end{abstract}


\section{Introduction}

The process of hydrogen reionisation is thought to have started with
ionised (H\,{\sevensize II}) regions around the first galaxies, which
later grew to surround groups of galaxies. Reionisation completed once
these H\,{\sevensize II} regions overlapped and occupied most of the
volume between galaxies. Much recent theoretical attention has focused
on the power spectrum (PS) of 21-cm emission from neutral hydrogen
(H\,{\sevensize I}) during the reionisation era (Zaldarriaga,
Furlanetto \& Hernquist~2004; Furlanetto et al.~2004; Morales et
al.~2005; Bowman et al.~2006). In particular, the ionisation structure
of the intergalactic medium (IGM) owing to UV emission associated with
star formation has been studied in detail using analytic~(Furlanetto
et al.~2004; Barkana~2007), numerical~(McQuinn et al.~2006; Iliev et
al.~2008), and more recently, semi-numerical models (Zahn et al.~2007;
Mesinger \& Furlanetto~2007; Geil \& Wyithe~2008).

These studies describe a scenario in which very large H\,{\sevensize
  II} regions form around clustered sources within overdense regions
of the IGM. The formation of these H\,{\sevensize II} regions has a
significant effect on the shape of the 21-cm PS because information on
the small scale features in the density field is erased from the
signal originating within the ionised regions. Conversely, the
creation of large ionised regions imprints large scale features on the
distribution of 21-cm intensity. The sum of these effects is to move
power from small to large scales, leaving a shoulder shaped feature on
the PS at the characteristic scale of the H\,{\sevensize II} regions
(Furlanetto et al.~2004). The detailed morphology of the ionisation
structure will therefore yield information about both the ionising
sources and the structure of absorbers in the IGM on small
scales~(McQuinn et al.~2006).  As reionisation leaves a strong imprint
on the 21-cm PS, measuring the latter has become a key goal for
learning about the reionisation epoch (Furlanetto, Oh \& Briggs~2006,
and references therein). In addition, since reionisation is driven by
galaxy formation, whose statistics reflect those of the underlying
density field, the detection of the redshifted 21-cm signal will not
only probe the astrophysics of reionisation, but also the matter PS
during the epoch of reionisation (McQuinn et al.~2006; Bowman, Morales
\& Hewitt~2007).  Finally, it has been recognised that
cross-correlating galaxy surveys with 21-cm maps could yield powerful
new insights into the morphology of reionisation, as well as eliminate
some of the difficulties related to foreground removal (Furlanetto \&
Lidz 2007; Wyithe \& Loeb 2007; Lidz et al 2008).  With these ideas as
motivation, several experiments are currently under development that
aim to detect the 21-cm signal during reionisation, including the Low
Frequency Array\footnote{http://www.lofar.org/} (LOFAR), the Murchison
Widefield Array\footnote{http://www.haystack.mit.edu/ast/arrays/mwa/}
(MWA) and the Precision Array to Probe Epoch of
Reionization\footnote{http://astro.berkeley.edu/$\sim$dbacker/eor/}
(PAPER), and more ambitious designs are being planned such as the
Square Kilometer Array\footnote{http://www.skatelescope.org/} (SKA).

Up until now only the component of H\,{\sevensize I} residing in the
IGM has been considered in relation to forecasts of the statistical
21-cm signal. However, after the completion of reionisation there is
known to be a residual H\,{\sevensize I} fraction of a few percent in
high density clumps which are believed to reside within galaxies
(e.g., Prochaska et al 2005).  This high density contribution to the
H\,{\sevensize I} content of the Universe is also present during the
reionisation era (we refer to this high density H\,{\sevensize I} as
\textit{galactic H\,{\sevensize I}} throughout the paper). Moreover,
because galaxies at high redshift are biased relative to the density
field, this galactic H\,{\sevensize I} contribution could provide a
significant perturbation to the predicted statistics of 21-cm
fluctuations.

Wyithe \& Loeb~(2007) have modelled the density dependent reionisation
process using a semi-analytic model that incorporates the important
physical processes associated with galaxy bias and radiative
feedback. In agreement with numerical simulations~(McQuinn et
al.~2006; Iliev et al.~2008), this model demonstrates that galaxy bias
leads to enhanced reionisation in overdense regions. In this paper we
use this model and its semi-numerical extension to explore the effect
that the H\,{\sevensize I} content of the biased galactic sources of
reionisation would have on the fluctuations in redshifted 21-cm
emission. We begin by describing our density dependent model of
reionisation in \S~\ref{model}. We then summarise the observed
evolution of the H\,{\sevensize I} density in \S~\ref{HI}, and our
results for the evolution of the 21-cm signal in \S~\ref{modelev}. We
next describe the effect of galactic H\,{\sevensize I} on the 21-cm
signal using analytic and semi-numerical models in
\S~\ref{model1}-\ref{seminum}, and discuss prospects for detection in
\S~\ref{noise}. We conclude in \S~\ref{conclusion}. Throughout this
paper we adopt a concordance cosmology for a flat $\Lambda$CDM
universe, $(\Omega_{\rm m},\Omega_{\Lambda},\Omega_{\rm
  b},h,\sigma_8,n) = (0.27,0.73,0.046,0.7,0.8,1)$, consistent with the
constraints from Komatsu et al.~(2008). All distances are in comoving
units unless stated otherwise.

\section{Density Dependent analytic model of reionisation}
\label{model}

In regions of the IGM that are overdense, galaxies will be
over-abundant for two reasons: first because there is more material
per unit volume to make galaxies, and second because small scale
fluctuations need to be of lower amplitude to form a galaxy when
embedded in a larger scale overdensity (the so-called {\it galaxy
  bias}; see Mo \& White~1996). Regarding reionisation of the IGM, the
first effect will result in a larger density of ionising
sources. However this larger density will be compensated by the
increased density of gas to be ionised, which also increases the
recombination rate.  The process of reionisation also contains several
layers of feedback.  Radiative feedback heats the IGM and results in
the suppression of low-mass galaxy formation (Efstathiou, 1992; Thoul
\& Weinberg~1996; Quinn et al.~1996; Dijkstra et al.~2004).  Such
feedback effects can potentially be more intense in overdense regions,
leading to weaker galaxy formation bias than might be expected from
simple linear bias models (Kramer et al 2006). Probing the morphology
of reionisation could potentially lead to constraints on such feedback
effects.

To compute the relation between the local overdensity and the
brightness temperature of redshifted 21-cm emission we use the model
described in Wyithe \& Loeb~(2007). Here we summarise the main
features of the model, and describe the additions made for the purpose
of including the possible contribution from galaxies.  The evolution
of the ionisation fraction by mass $Q_{\delta,R}$ of a particular
region of IGM with scale $R$ and overdensity $\delta$ (at observed
redshift $z_{\rm obs}$) may be written as
\begin{eqnarray}
\label{history}
\nonumber
\frac{dQ_{\delta,R}}{dt} &=& \frac{N_{\rm ion}}{0.76}\left[Q_{\delta,R} \frac{dF_{\rm col}(\delta,R,z,M_{\rm ion})}{dt} \right.\\
\nonumber
&&\hspace{5mm}+ \left.\left(1-Q_{\delta,R}\right)\frac{dF_{\rm col}(\delta,R,z,M_{\rm min})}{dt}\right]\\
&-&\alpha_{\rm B}Cn_{\rm H}^0\left(1+\delta\frac{D(z)}{D(z_{\rm obs})}\right) \left(1+z\right)^3Q_{\delta,R},
\end{eqnarray}
where $N_{\rm ion}$ is the number of photons entering the IGM per
baryon in galaxies, $\alpha_{\rm B}$ is the case-B recombination
coefficient, $C=2$ is the clumping factor (which we assume, for
simplicity, to be constant), and $D(z)$ is the growth factor between
redshift $z$ and the present time. The production rate of ionising
photons in neutral regions is assumed to be proportional to the
collapsed fraction $F_{\rm col}$ of mass in halos above the minimum
thresholds in neutral ($M_{\rm min}$), and in ionised ($M_{\rm ion}$)
regions. We assume $M_{\rm min}$ to correspond to a virial temperature
of $10^4$K, representing the hydrogen cooling threshold, and $M_{\rm
  ion}$ to correspond to a virial temperature of $10^5$K, representing
the mass below which infall is suppressed from an ionised IGM
(Dijkstra et al.~2004). In a region of co-moving radius $R$ and mean
overdensity $\delta(z)=\delta D(z)/D(z_{\rm obs})$ [specified at
redshift $z$ instead of the usual $z=0$], the relevant collapsed
fraction is obtained from the extended Press-Schechter~(1974) model
(Bond et al.~1991) as
\begin{equation}
F_{\rm col}(\delta,R,z) = \mbox{erfc}{\left(\frac{\delta_{\rm
c}-\delta(z)}{\sqrt{2\left(\left[\sigma_{\rm
gal}\right]^2-\left[\sigma(R)\right]^2\right)}}\right)},
\end{equation}
where $\mbox{erfc}(x)$ is the complementary error function,
$\sigma^2(R)$ is the variance of the density field smoothed on a scale
$R$, and $\sigma^2_{\rm gal}$ is the variance of the density field
smoothed on a scale $R_{\rm gal}$, corresponding to a mass scale of
$M_{\rm min}$ or $M_{\rm ion}$ (both evaluated at redshift $z$ rather
than at $z=0$).  In this expression, the critical linear overdensity
for the collapse of a spherical top-hat density perturbation is
$\delta_c\approx 1.69$.

The model assumes that on large (linear regime) scales most ionising
photons are absorbed locally, so that the ionisation of a region is
caused by nearby ionisation sources. This assumption is certainly
justified during the early stages of reionisation, when the mean free
path for ionising photons is short.  However, even later in the
reionisation process, the mean free path always remains smaller than
the characteristic H\,{\sevensize II} bubble size (it could be smaller
if mini-halos or pockets of residual H\,{\sevensize I} block ionising
photons between the sources and the edge of the H\,{\sevensize II}
region.) Our local ionisation assumption is therefore valid as long as
the characteristic bubble size is smaller than the spatial scale of
the correlations we consider. This requirement is met in regimes where
the fraction of regions at a particular scale are fully ionised is
very low. This point is discussed quantitatively in Wyithe \&
Morales~(2007).

\section{The H\,{\sevensize I} content of galaxies}
\label{HI}

\begin{figure*}
\includegraphics[width=13.cm]{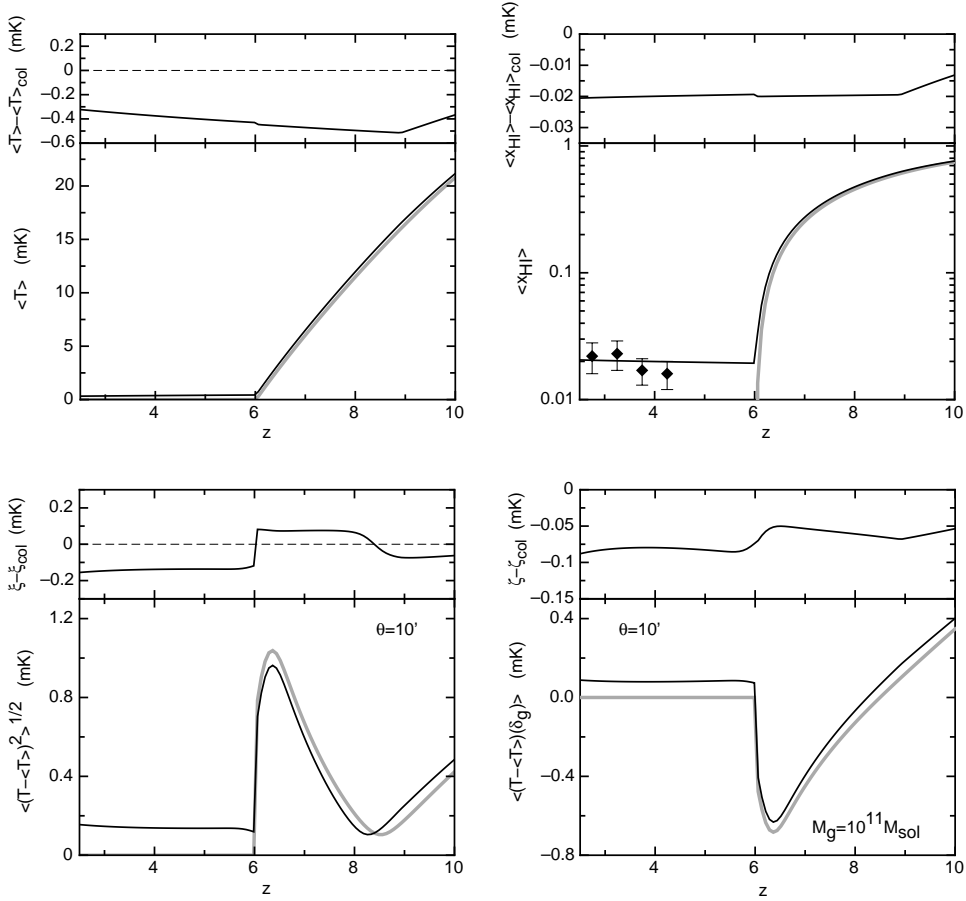} 
\caption{Model-I associating galactic H\,{\sevensize I} with star forming galaxies. \textit{Upper Left:} The evolution of the mean brightness temperature with redshift. \textit{Upper Right:}  The evolution of the mean H\,{\sevensize I} fraction with redshift. The observed points are inferred from observations of damped Ly$\alpha$ absorbers (Prochaska et al.~2005). \textit{Lower Left:} The evolution of the auto-correlation function of brightness temperature with redshift (computed within regions of size $10'$). \textit{Lower Right:} The evolution of the cross-correlation function with redshift (computed within regions of size $10'$). In each case the grey and dark lines correspond to models which exclude and include the galactic H\,{\sevensize I} respectively. The small upper sections of each panel show the difference between the models without galactic H\,{\sevensize I} and including galactic H\,{\sevensize I}. }
\label{fig1}
\end{figure*} 
Following the completion of reionisation, around 2\% of the hydrogen
content of the Universe is observed to be in H\,{\sevensize I} within
galaxies. Moreover, this two percent fraction is observed to be
roughly constant from $z\sim1$ to $z\sim5$ (e.g. Prochaska et
al.~2005). However, to estimate the effect of galactic H\,{\sevensize
  I} on 21-cm fluctuation statistics during the reionisation epoch we
must extrapolate to higher redshift. There are two competing
factors. On the one hand the collapsed fraction of mass rises as
reionisation progresses, which would imply a galactic H\,{\sevensize
  I} fraction that decreases towards high redshift. On the other hand,
the fact that the H\,{\sevensize I} fraction is constant below
$z\sim5$ suggests that the fraction of hydrogen within galaxies that
is in H\,{\sevensize I} increases with redshift. In this paper we
conservatively assume that the fraction of gas within galaxies that is
H\,{\sevensize I} increases at a rate that preserves the 2\% galactic
H\,{\sevensize I} mass fraction (relative to the total hydrogen
content of the Universe) into the reionisation epoch. At redshifts
where there is not a sufficient collapsed fraction to maintain the 2\%
value, the galactic H\,{\sevensize I} fraction is allowed to drop
below 2\%.

This is obviously a fairly simplistic model. The actual physics of the
HI content in galaxies depends both on star formation and feedback
processes within the galaxy, as well as on the intergalactic UV
background. For instance, as the UV background falls toward high
redshift, the HI fraction within galaxies could increase, and perhaps
conspires to cancel with the decreasing collapse fraction to maintain
the observed roughly constant H\,{\sevensize I} fraction of $\sim
2\%$, as seen from $z\sim 1-5$. Indeed, the fraction in protogalactic
H\,{\sevensize I} may be even larger than the $\sim 2\%$ we assume in
the pre-reionisation epoch, particularly if there is a significant
population of minihalos\footnote{Since these are photoevaporated as
  reionisation progresses (Shapiro et al 2004), they will reside
  primarily in neutral regions, and present an additional biased
  H\,{\sevensize I} contribution that is anti-correlated with galaxies
  but which we do not model here.} with $T_{\rm vir} <
10^{4}$K. Modelling this quantity is difficult and beyond the scope of
this paper. For now, we note that our empirical extrapolation becomes
increasingly uncertain at high redshift; it is probably reasonable in
the redshift range $z\sim 6-8$, where some of the most interesting
effects (such as the change of sign of the cross-correlation between
galaxies and 21-cm emission) take place.

\section{Evolution of the 21-cm intensity signal}
\label{modelev}

To illustrate the contribution of galactic H\,{\sevensize I}, we use
the following example. We first find the value of $N_{\rm ion}$ that
yields overlap of ionised regions at the mean density IGM by $z\sim6$
(Fan et al.~2006; Gnedin \& Fan~2006; White et al.~2003), and then
integrate equation~(\ref{history}) as a function of $\delta$ and $R$.
At a specified redshift, this yields the filling fraction of ionised
regions within the IGM on various scales $R$ as a function of
overdensity. We may then calculate the corresponding 21-cm brightness
temperature contrast owing to H\,{\sevensize I} in both the IGM and
galaxies
\begin{eqnarray}
\nonumber
\label{eqT}
T(\delta,R) &=& 22\mbox{mK}\left(\frac{1+z}{7.5}\right)^{1/2}\\
&&\hspace{-5mm}\times\left(1-Q_{\delta,R}-Q^{\rm col}_{\delta,R}\right)\left(1+\frac{4}{3}\delta\right),
\end{eqnarray}
where the pre-factor of 4/3 on the overdensity refers to the
spherically averaged enhancement of the brightness temperature due to
peculiar velocities in overdense regions (Bharadwaj \& Ali~2005;
Barkana \& Loeb~2005), and $1-Q^{\rm col}_{\delta,R}$ is the mass
averaged fraction of H\,{\sevensize I} in galaxies relative to the
total hydrogen content of the Universe.  Note that we only account for
brightness temperature fluctuations due to variation in the density
and ionisation fraction of the IGM. We ignore variations in the spin
temperature (which we assume to be always much greater than the CMB
temperature) due to fluctuations in heating and radiative Ly$\alpha$
coupling; these are important primarily early in the reionisation
process, when galaxies are in any case rare.

\subsection{The auto-correlation function of 21-cm intensity fluctuations}

Since the underlying probability distribution of overdensities
$dP/d\delta$ is known (a Gaussian of variance $\sigma D$), we may
compute the observed probability distribution for $T$,
\begin{equation}
\frac{dP}{dT}(\theta) \propto \frac{dP}{d\delta}\left|\frac{\partial\delta}{\partial T}\right|.
\end{equation}
The second moment of this distribution corresponds to the
auto-correlation function of brightness temperature smoothed on an
angular radius $\theta$:
\begin{eqnarray}
\nonumber
\xi\equiv\langle \left(T-\langle T\rangle\right)^{2}\rangle &=&\\
 &&\hspace{-20mm}\left[\frac{1}{\sqrt{2\pi}\sigma(R)}\int d\delta \left(T(\delta,R)-\langle T\rangle\right)^2e^{-\frac{\delta^2}{2\sigma(R)^2}} \right],
\end{eqnarray}
where 
\begin{equation}
\langle T\rangle = \frac{1}{\sqrt{2\pi}\sigma(R)}\int d\delta~ T(\delta,R)e^{-\frac{\delta^2}{2\sigma(R)^2}} ,
\end{equation}
and $\sigma(R)$ is the variance of the density field (at redshift $z$)
smoothed on a scale $R$.

\subsection{The cross-correlation between galaxies and 21-cm intensity fluctuations}
\label{cc_bn_gand21}
The inclusion of galactic H\,{\sevensize I} in the calculation of the
21-cm signal will also modify predictions for the cross-correlation of
21-cm emission with galaxies.  The cross-correlation function of
galaxy overdensity and brightness temperature smoothed on an angular
radius $\theta$ is
\begin{eqnarray}
\label{Tvar}
\nonumber
\zeta\equiv\langle \left(T-\langle T\rangle\right)\delta_{\rm g}\rangle &=&\\
 &&\hspace{-25mm}\left[\frac{1}{\sqrt{2\pi}\sigma(R)}\int d\delta \left(T(\delta,R)-\langle T\rangle\right)\times \frac{4}{3}\delta_{\rm g} e^{-\frac{\delta^2}{2\sigma(R)^2}} \right],
\end{eqnarray}
where $\delta_{\rm g} = b\delta$, and $b(M_{\rm g},z)$ is the galaxy
bias for a halo of mass $M_{\rm g}$ at redshift $z$ (Sheth, Mo \&
Tormen~2001). To calculate the cross-correlation function we assume
$M_{\rm g}=10^{11}M_\odot$ for the host mass of observed galaxies
throughout this paper. This host mass yields a density of galaxies
that is comparable to the density observed in the Subaru Deep Fields
(Kashikawa et al.~2006) at $z\sim6.6$ (Furlanetto \& Lidz~2007).

\begin{figure*}
\includegraphics[width=13.cm]{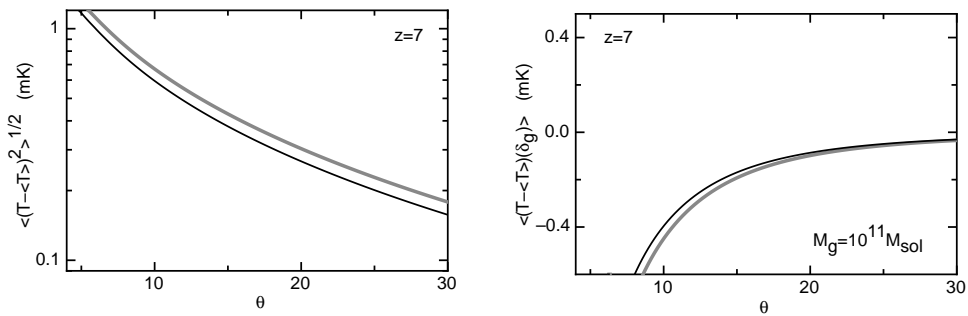} 
\caption{Model-I associating galactic H\,{\sevensize I} with star forming galaxies. \textit{Left:} The angular dependence of the auto-correlation function of brightness temperature (computed at $z=7$). \textit{Right:} The angular dependence of the cross-correlation function (computed at $z=7$). In each case the grey and dark lines correspond to models which exclude and include the galactic H\,{\sevensize I} respectively. }
\label{fig2}
\end{figure*} 
\section{model-I: galactic H\,{\sevensize I} associated with star forming galaxies}
\label{model1}

In this section we use our analytic model to estimate the effect of
galactic H\,{\sevensize I} on 21-cm fluctuations. We assume that the
star forming galaxies responsible for reionisation are the harbourers
of galactic H\,{\sevensize I}, and calculate the fraction of galactic
hydrogen that is in H\,{\sevensize I} form
\begin{eqnarray}
\nonumber 
f_{\rm H{\scriptscriptstyle I},col} &=& \\
&&\hspace{-15mm}\min\left(1,\frac{0.02}{F_{\rm col}(z,M_{\rm min})\left(1-Q\right) + F_{\rm col}(z,M_{\rm ion})Q}\right),
\end{eqnarray}
where $F_{\rm col}$ and $Q$ are computed for the mean universe
($\delta=0$, $R=\infty$).  This enforces the condition that the
fraction of HI in galaxies is either $\sim 2\%$, or equal to the
(lower) collapse fraction in galaxies.  We can then compute the
fraction of cosmic hydrogen that is galactic H\,{\sevensize I} in
regions of overdensity $\delta$ and radius $R$
\begin{eqnarray}
\label{xHI}
\nonumber
1-Q^{\rm col}_{\delta,R} &=& f_{\rm H{\scriptscriptstyle I},col} \left[ Q_{\delta,R} F_{\rm col}(\delta,R,z,M_{\rm ion}) \right.\\
&&\hspace{10mm} + \left.\left(1-Q_{\delta,R}\right)F_{\rm col}(\delta,R,z,M_{\rm min})\right]. 
\end{eqnarray}
This contribution to the H\,{\sevensize I} fraction of the Universe
can be used with equation~(\ref{eqT}) to estimate the effect of
galactic H\,{\sevensize I} on the statistics of 21-cm emission.

The evolution of mean brightness temperature, together with the mean
H\,{\sevensize I} fraction $\langle x_{\rm H{\scriptscriptstyle I}}
\rangle=1-Q-Q^{\rm col}$ are plotted in Figure~\ref{fig1}. The data
points show the observed fraction from damped Ly$\alpha$ absorbers for
comparison (Prochaska et al~2005). The solid dark lines show the case
including the galactic H\,{\sevensize I} contribution. The thick grey
line shows the IGM-only model for comparison. The residuals plotted
above each panel show the magnitude of the effect when the galaxy
contribution is ignored.  The inclusion of galactic H\,{\sevensize I}
increases the global 21-cm signal slightly as expected.

Examples of auto-correlation functions are shown in Figures~\ref{fig1}
and \ref{fig2}. The auto-correlation function is plotted as a function
of redshift at $\theta=10^\prime$ (Figure~\ref{fig1}), and as a
function of $\theta$ at $z=7$ (Figure~\ref{fig2}). In
Figure~\ref{fig1} the residuals are plotted above each panel to show
the magnitude of the error introduced when the galaxy contribution is
ignored.  Early in the reionisation process, before the appearance of
H\,{\sevensize II} regions begins to dominate the fluctuation
amplitude, the inclusion of a galactic fraction enhances the 21-cm
fluctuations. However, in contrast to the mean signal, the size of
fluctuations are reduced by the presence of galaxies late in the
reionisation era. This reduction can be traced to the fact that
galaxies are biased towards overdense regions, while neutral IGM is
biased towards underdense regions which are reionised last. As a
result, the inclusion of a galactic H\,{\sevensize I} fraction reduces
the intensity contrast between overdense and underdense regions. The
size of the residuals shows that within this model, the 2\% galactic
H\,{\sevensize I} fraction reduces the fluctuation amplitude over a
large fraction of the reionisation epoch. At $z\sim7$, it reduces the
fluctuation amplitude by around 10\% at all angles (which makes sense,
since galactic HI constitutes about $\sim 10\%$ of the total HI
content of the universe at that point).

Examples of the cross-correlation functions are shown in
Figure~\ref{fig1} as a function of redshift at fixed angle
($\theta=10'$), and in Figure~\ref{fig2} as a function of angle at
fixed redshift ($z=7$). The inclusion of galactic H\,{\sevensize I}
reduces the amplitude of the cross-correlation function by about 10\%
(relative to the case where galactic H\,{\sevensize I} is ignored)
late in the reionisation era for the reasons discussed in
\S~\ref{cc_bn_gand21}. In addition, the cross-correlation changes sign
at overlap because the H\,{\sevensize I} content of the Universe
shifts from being dominated by the underdense regions of IGM to
H\,{\sevensize I} in galaxies (which reside in overdense regions).
This sign change will provide an unambiguous pointer to the redshift
at which reionisation ends. Moreover, because the sign change will
occur over a narrow frequency interval, it should provide an important
check on possible sources of systematic error that could be present in
measurements of the PS of intensity fluctuations which is always
positive by construction.

\section{Model-II: Varying host mass for galactic H\,{\sevensize I}}
\label{fixedM}

\begin{figure*}
\includegraphics[width=13.cm]{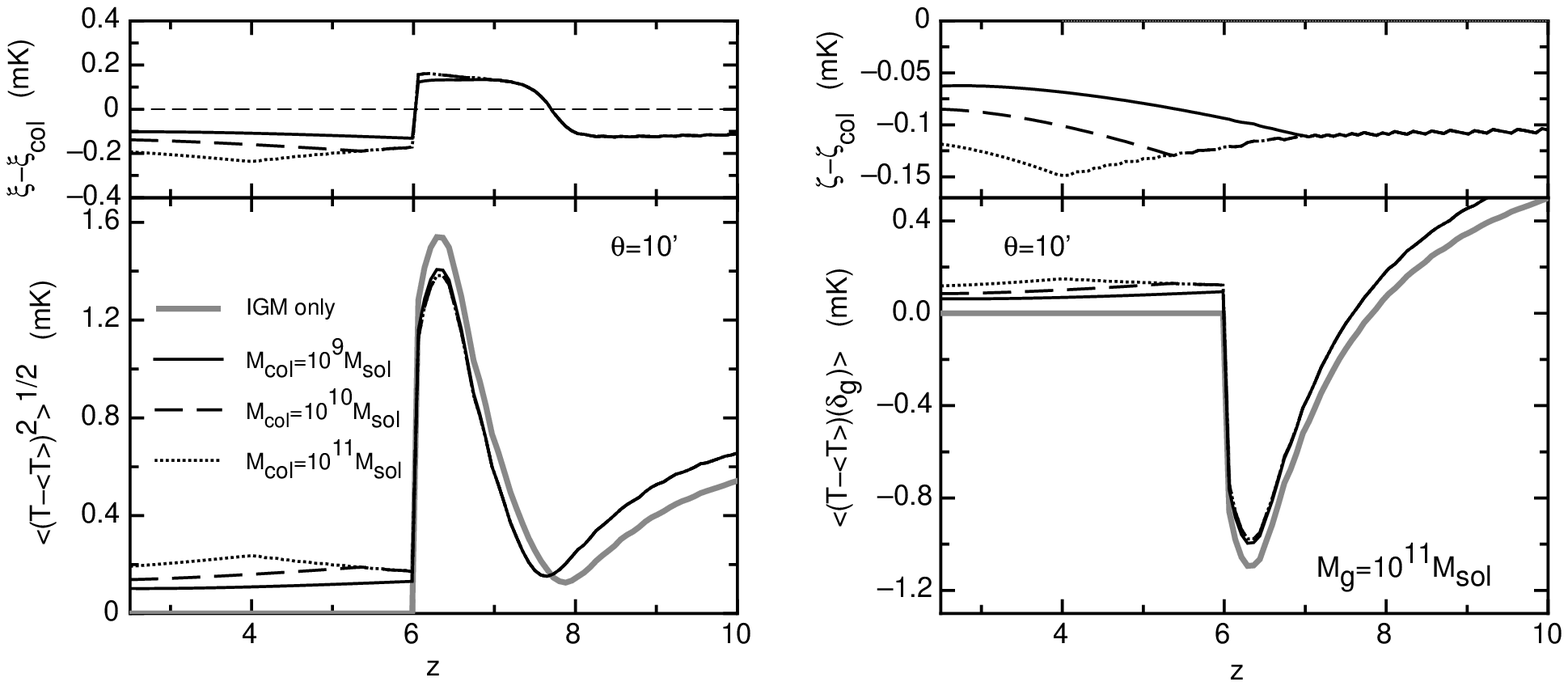} 
\caption{Model-II with fixed characteristic host mass for galactic H\,{\sevensize I}. \textit{Left:} The evolution of the auto-correlation function of brightness temperature with redshift (computed within regions of size $10'$). \textit{Right:} The evolution of the cross-correlation function with redshift (computed within regions of size $10'$). In each case the grey and dark lines correspond to models which exclude and include galactic H\,{\sevensize I} respectively. The small upper section of each panel shows the difference between the models without and including galactic H\,{\sevensize I}. }
\label{fig3}
\vspace{7mm}
\includegraphics[width=13.cm,angle=0]{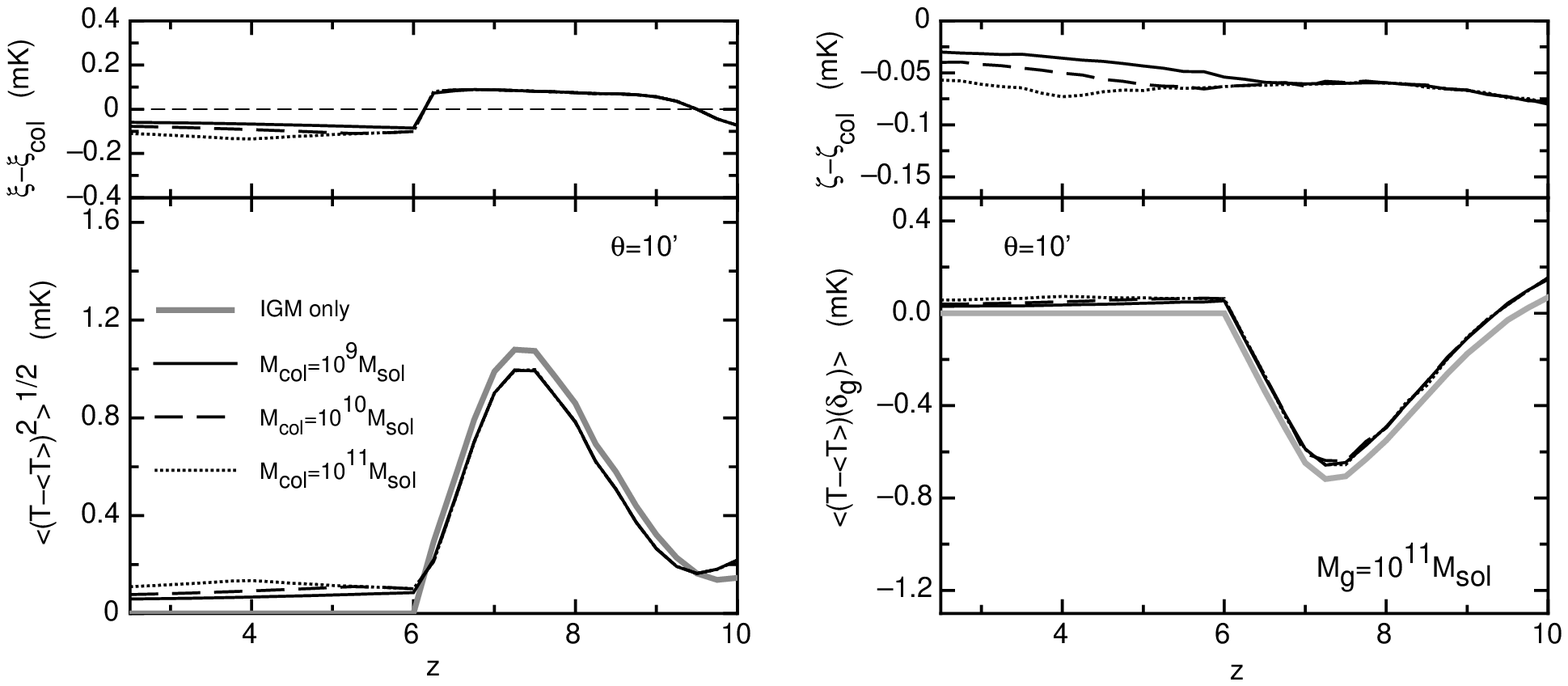} 
\caption{As per Figure~\ref{fig3}, but computed using the semi-numeric rather than analytic model.}
\label{fig4}
\end{figure*}

The calculations presented in Figures~\ref{fig1} and \ref{fig2} assume
that the galactic H\,{\sevensize I} contributing to the 21-cm emission
is found in the star forming systems. However it is possible that the
connection between galactic H\,{\sevensize I} and star formation is
not direct, either because there is H\,{\sevensize I} remaining in
older low mass galaxies, or because the fraction of galactic gas that
is H\,{\sevensize I} is host mass dependent. As presented, our model
is not able to address either of these issues. Therefore, to ascertain
the possible range of influence that galactic H\,{\sevensize I} might
have on the 21-cm fluctuation statistics, we have assumed a
characteristic mass $M_{\rm col}$ for the hosts of galactic
H\,{\sevensize I}, and replaced the expression for $T_{\rm IGM}$
(equation~\ref{eqT}) with
\begin{eqnarray}
\nonumber
\label{TIGM}
T_{\rm IGM}(\delta,R) &=& 22\mbox{mK}\left(\frac{1+z}{7.5}\right)^{1/2}\\
&&\hspace{-13mm}\times\left[(1-Q_{\delta,R})\left(1+\frac{4}{3}\delta\right) + x_{\rm H{\scriptscriptstyle I},col}\left(1+\frac{4}{3}b\delta\right)\right],
\end{eqnarray}
where we have defined the mass averaged fraction of cosmic hydrogen
that is galactic H\,{\sevensize I} ($x_{\rm H{\scriptscriptstyle
    I},col}$).  We note that in order to maintain self consistency,
the host mass of the galactic H\,{\sevensize I} must correspond to a
collapsed fraction [$F_{\rm col}(M_{\rm col})]$ that is in excess of
$x_{\rm H{\scriptscriptstyle I},col}$. At a particular redshift the
maximum possible host mass ($M_{\rm col,x}$) corresponding to a
collapsed fraction that equals $x_{\rm H{\scriptscriptstyle I},col}$
is given by evaluation of
\begin{equation}
F_{\rm col}(\delta,R,z,M_{\rm col,x})=x_{\rm H{\scriptscriptstyle I},col}.
\end{equation}
The host mass at which the galaxy bias ($b$)  in
equation~(\ref{TIGM}) is evaluated is therefore
\begin{equation}
\label{minmass}
M=\min\left(M_{\rm col,x},M_{\rm col}\right).
\end{equation}
With these modifications for the galactic H\,{\sevensize I}
distribution, we may compute the 21-cm fluctuation statistics as
before using equation~(\ref{Tvar}).

The resulting auto-correlation functions are shown in
Figure~\ref{fig3} as a function of redshift at fixed angle
($\theta=10'$). Three values for $M_{\rm col}$ are shown,
$10^9M_\odot$, $10^{10}M_\odot$ and $10^{11}M_\odot$.  Late in the
reionisation era, the amplitude of the auto-correlation function is
decreased by $\sim10\%$ (relative to the case where galactic
H\,{\sevensize I} is ignored). Below $z\sim6$ larger, more biased
hosts of galactic H\,{\sevensize I} result in a more significant
increase of the fluctuation amplitude. However at $z\ga6$ $M_{\rm
  col,x}<M_{\rm c}$, and hence our model predicts the same evolution
for different values of $M_{\rm col}$ (see
equation~\ref{minmass}). The cross-correlation function is also shown
in Figure~\ref{fig3}. The cross-correlation changes sign at overlap as
before. More massive hosts of the galactic H\,{\sevensize I} result in
a cross-correlation with a larger amplitude following reionisation.

\section{Semi-Numerical simulations}
\label{seminum}

The analytic calculations described thus far provide a qualitative
description of the effect of galactic H\,{\sevensize I} on the
statistics of 21-cm fluctuations and their association with
galaxies. However, this analytic model is unable to describe
fluctuations on scales comparable to the characteristic bubble scale
once reionisation becomes established (Wyithe \& Morales 2007). In the
next section of this paper we therefore investigate the statistics of
21-cm fluctuations including galactic H\,{\sevensize I} in a
semi-numerical model for the reionisation of a three-dimensional
volume of the IGM (Geil \& Wyithe~2008; Zahn et al.~2007; Mesinger \&
Furlanetto~2007).  Our semi-numerical simulations are based on
model-II with respect to the galactic H\,{\sevensize I} contribution.

\subsection{The ionisation field}
\label{ionisationfield}

Our modelling follows the procedure outlined in Geil \& Wyithe~(2008),
and we refer the reader to that paper for details of the model. The
model employs a semi-analytic prescription for the reionisation
process, which is combined with a realisation of the density field.
We construct an ionisation field based on a Gaussian random field for
the overdensity of mass, combined with the value of the ionised
fraction $Q_{\delta,R}$ (equation~\ref{history}) as a function of
overdensity $\delta$ and smoothing scale $R$.  We repeatedly filter
the linear density field at logarithmic intervals on scales comparable
to the box size down to the grid scale size. For all filter scales,
the ionisation state of each grid position is determined using
$Q_{\delta,R}$ and deemed to be fully ionised if
$Q_{\delta,R}\geqslant 1$. All voxels within a sphere of radius $R$
centred on these positions are flagged and assigned $Q_{\delta,R}=1$,
while the remaining non-ionised voxels are assigned an ionised
fraction of $Q_{\delta,R_{\textrm{f,min}}}$, where
$R_{\textrm{f,min}}$ corresponds to the smallest smoothing scale. A
voxel forms part of an H\,{\sevensize II} region if $Q_{\delta,R}>1$
on any scale $R$. In this paper we present simulations corresponding
to a linear density field of resolution $256^{3}$, with a comoving
side length of 512\,Mpc. This procedure yields an ionisation map
$Q_{\rm ion}({\textit{\bf x}})$ as a function of position
${\textit{\bf x}}$.

\begin{figure*}
\includegraphics[width=13cm,angle=0]{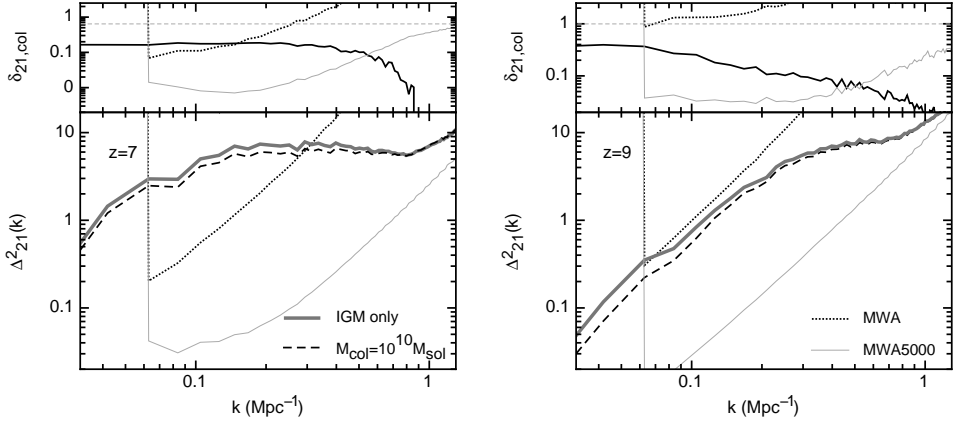} 
\caption{{\em Lower} panels: 21-cm power spectra computed using semi-numerical models at $z=7$ (left) and $z=9$ (right). In each case models are shown for a host mass of $M_{\rm col}=10^{10}M_\odot$ for the galactic H\,{\sevensize I}, as well as for the IGM-only case. {\em Upper} panels: The relative perturbation on the PS due to galactic H\,{\sevensize I} [$\delta_{\rm 21,col}=(\Delta^2_{\rm 21}-\Delta^2_{\rm 21,col})/\Delta^2_{21}$]. Also plotted for comparison is the estimated sensitivity (lower panels), and the sensitivity relative to $\Delta^2_{\rm 21}$ (upper panels) within bins of width $\Delta k=k/10$ assuming 1000 hr integration on 1 field for the MWA and MWA5000 (dotted and thin-grey lines).}
\label{fig5}
\end{figure*} 

Having computed the ionisation field we then find the distribution of
the galactic H\,{\sevensize I} component. Rather than try to assign
individual galaxies to the simulation we make the approximation that
Poisson noise will be negligible and assign a smooth density of
H\,{\sevensize I}
\begin{equation}
\rho_{\rm HI,col}(\vec{x})=x_{\rm HI,col}\rho_{H}[1+b\delta(\vec{x})],
\end{equation}
where $\rho_{\rm HI,col}$ and $\rho_{\rm H}$ are the mass averaged
densities of H\,{\sevensize I} that is collapsed in galaxies, and the
total hydrogen density respectively. The position dependent brightness
temperature of the simulation box becomes
\begin{eqnarray}
\label{eqT}
\nonumber
T_{\rm IGM}(\vec{x}) &=& 22\mbox{mK}\left(\frac{1+z}{7.5}\right)^{1/2}\\
&&\hspace{0mm}\times\left[(1-Q_{\rm ion})\left(1+\delta\right) + x_{\rm HI,col}(1+b\delta)\right].
\end{eqnarray}
Note that our semi-numerical model does not compute peculiar
velocities, and so equation~(\ref{eqT}) does not include a peculiar
velocity induced enhancement of the brightness temperature in
overdense regions.  As in earlier sections we consider a model in
which the mean IGM is reionised at $z=6$. We again assume that star
formation proceeds in halos above the hydrogen cooling threshold in
neutral regions of IGM. In ionised regions of the IGM star formation
is assumed to be suppressed by radiative feedback (see
\S\,\ref{model}).

\subsection{Variance in 21-cm emission} 

Since our semi-numerical model computes the three dimensional
ionisation structure in the IGM, including the effect of
H\,{\sevensize II} regions, we can use it to compute the evolution of
the variance. The resulting auto-correlation functions are shown in
the left hand panel of Figure~\ref{fig4} as a function of redshift at
fixed angle ($\theta=10'$). As in the analytic model presented in
Figure~\ref{fig3}, three values for $M_{\rm col}$ are shown, $10^9
M_{\odot}$, $10^{10}M_{\odot}$ and $10^{11} M_{\odot}$. These curves
can be compared directly with the analytic approximation. This
comparison shows that while the analytic model yields the correct
qualitative behaviour, it is not quantitatively correct, both in terms
of the redshift and amplitude of maximum fluctuations. However, in
agreement with the analytic calculation, our semi-numerical
calculations of the auto-correlation function show that the
fluctuation amplitude is modified at the level of 10-20\% (relative to
the case where galactic H\,{\sevensize I} is ignored) by the presence
of a 2\% galactic H\,{\sevensize I} fraction throughout the
reionisation era. The semi-numerical calculations of the
cross-correlation function are also shown in the right hand panel of
Figure~\ref{fig4} as a function of redshift at fixed angle
($\theta=10'$). Comparison with Figure~\ref{fig3} also shows that the
analytic calculation of the cross-correlation is qualitatively
correct, but does not predict the correct quantitative evolution.

\subsection{21-cm power spectrum}

Figure~\ref{fig5} shows 21-cm PS computed from the semi-numerical
simulations at $z=7$ (\textit{left}) and $z=9$ (\textit{right}). We
plot the dimensionless PS $\Delta^{2}_{21}(k)=k^3/(2\pi^2)P_{21}(k)$,
where $P_{21}$ is the PS of 21-cm fluctuations. Both the case of the
IGM alone ($\Delta_{21}^{2}$), and the case assuming a host mass of
$M_{\rm col}=10^{10}M_\odot$ for the galactic H\,{\sevensize I}
($\Delta^{2}_{\rm 21,col}$) are shown (lower panels). We also show the
relative fluctuation $\delta_{\rm 21,col}=(\Delta^2_{\rm
  21}-\Delta^2_{\rm 21,col})/\Delta^2_{21})$ of the 21-cm PS owing to
galactic H\,{\sevensize I} (upper panels).

These figures illustrate the shoulder in the PS that corresponds to
the typical bubble scale and which is due to the movement of power
from small to large scales that accompanies the formation of
H\,{\sevensize II} regions. Because the galactic H\,{\sevensize I} is
biased towards H\,{\sevensize II} regions, this movement of power is
lessened when the contribution of galactic H\,{\sevensize I} is
considered. Hence the PS at large scales evaluated from simulations
which include H\,{\sevensize I} in galaxies is lower than the PS
computed using the IGM alone. The effect of galactic H\,{\sevensize I}
is therefore to change the shape of the 21-cm PS rather than just the
amplitude.

\subsection{21cm-galaxy cross power spectrum}

\begin{figure*}
\includegraphics[width=13.5cm,angle=0]{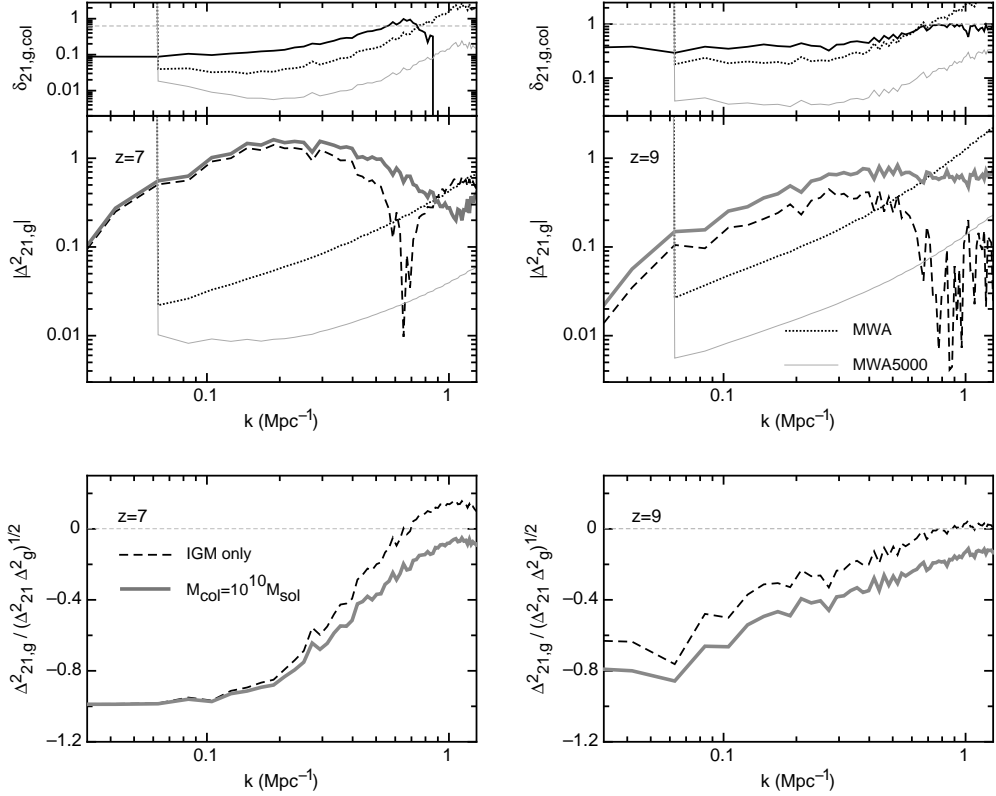} 
\caption{21cm-galaxy cross PS computed using semi-numerical models. {\em Central} and {\em Lower} panels: The modulus and coefficient of the 21cm-galaxy cross PS respectively computed at $z=7$ (left) and $z=9$ (right). In each case models are shown assuming a host mass of $M_{\rm col}=10^{10}M_\odot$  for the galactic H\,{\sevensize I}, as well as the IGM-only case. The host mass of observed galaxies is assumed to be $M_{\rm g}=10^{11}M_\odot$. {\em Upper} panels: The relative perturbation on the PS due to galactic H\,{\sevensize I} [$\delta_{\rm 21,g,col}=(\Delta^2_{\rm 21,g}-\Delta^2_{\rm 21,g,col})/\Delta^2_{\rm 21,g}$]. Also plotted for comparison is the estimated sensitivity (central panels), and the sensitivity relative to $\Delta^2_{\rm 21,g}$ (upper panels) within bins of width $\Delta k=k/10$ assuming 1000 hr integration on 1 field for the MWA and MWA5000 (dotted and thin-grey lines). To estimate this sensitivity galaxies were assumed to be observed down to $M_{\rm g}=10^{11}M_\odot$ over the entire field.}
\label{fig6}
\end{figure*}

The central and upper panels of Figure~\ref{fig6} show the modulus of
the dimensionless 21cm-galaxy cross PS 
with ($\Delta_{\rm 21,g,col}^{2}$) and without ($\Delta_{\rm 21,g}^{2}$) galactic H\,{\sevensize I} as well as the galactic H\,{\sevensize I} induced fluctuation [$\delta_{\rm 21,g,col}=(\Delta_{\rm 21,g}^{2}-\Delta_{\rm 21,g,col}^{2})/\Delta_{\rm 21,g}^{2}$], computed
from the semi-numerical simulations at $z=7$ (\textit{left}) and $z=9$
(\textit{right}). Again a host mass of $M_{\rm col}=10^{10}M_\odot$ is
considered for the galactic H\,{\sevensize I}. An observed galaxy mass of $M_{\rm g}=10^{12}M_\odot$ is assumed. In the lower panels of Figure~\ref{fig6} we show the corresponding coefficient of the 21cm-galaxy cross PS.

Since overdense regions, where galaxies are concentrated, are
reionised first these figures show an anti-correlation on large
scales, which drops in strength to zero on small scales after the
formation of H\,{\sevensize II} regions. We find that the inclusion of
galactic H\,{\sevensize I} lessens the amplitude of the\textsl{}
anti-correlation, since a fraction of H\,{\sevensize I} is now
co-located with the galaxies inside the H\,{\sevensize II} regions. In
addition, the cross PS changes sign on small scales which reflects the
correlation of galaxies with the galactic H\,{\sevensize I} inside the
H\,{\sevensize II} regions, where no power is contributed in 21-cm
fluctuations of the IGM. Thus, with the caveat that galaxies can be
selected without bias from IGM absorption (which may not be true if,
for instance, the galaxies are selected in Ly$\alpha$ emission---see
Lidz et al 2008 for discussion for the latter case), the scale at
which the 21cm-galaxy cross PS changes sign could be used to probe the
scale of H\,{\sevensize II} regions late in the reionisation era.

\section{Sensitivity to the effect of galactic H\,{\sevensize I} on 21-cm fluctuations}
\label{noise}

Before concluding, we compute the sensitivity with which the effect of
galactic H\,{\sevensize I} could be detected using forthcoming
low-frequency arrays.

\subsection{Sensitivity to the 21-cm PS}
To compute the sensitivity $\Delta P_{21}(\vec{k})$ of a
radio-interferometer to the 21-cm PS, we follow the procedure outlined
by McQuinn et al.~(2006) and Bowman, Morales \& Hewitt~(2007) [see
also ~Wyithe, Loeb \& Geil~(2008)]. The important issues are discussed
below, but the reader is referred to these papers for further
details. The sensitivity to the PS comprises components due to the
thermal noise, and due to sample variance within the finite volume of
the observations.  We consider observational parameters corresponding
to the design specifications of the MWA, and of a hypothetical
followup to the MWA (termed the MWA5000).  In particular the MWA is
assumed to comprise a phased array of 500 tiles. Each tile contains 16
cross-dipoles to yield an effective collecting area of $A_{\rm
  e}=16(\lambda^2/4)$ (the area is capped for $\lambda>2.1$m). The
physical area of a tile is $A_{\rm tile}=16$m$^2$.  The tiles are
distributed according to a radial antenna density of $\rho(r)\propto
r^{-2}$, within a diameter of 1.5km and outside of a flat density core
of radius 18m.  The MWA5000 is assumed to follow the basic design of
the MWA. The quantitative differences are that the telescope is
assumed to have 5000 tiles within a diameter of 2km, with a flat
density core of 80m. In each case we assume 1 field is observed for
1000 hr. Following the work of McQuinn et al.~(2006) we assume that
foregrounds can be removed over $8$MHz bins, within a bandpass of
$32$MHz [foreground removal therefore imposes a minimum on the
wave-number accessible of $k_{\rm
  min}\sim0.04[(1+z)/7.5]^{-1}$Mpc$^{-1}$].

The sensitivity to the 21-cm PS per mode may be written
\begin{equation}\label{thermal_noise}
\delta P_{21}(k,\theta) = \left[\frac{T_{\rm sky}^2}{\Delta\nu t_{\rm int}} \frac{D^2 \Delta D}{n(k_{\perp})}\left(\frac{\lambda^2}{A_e}\right)^2\right] + P_{21}(k,\theta),
\end{equation}
where $D$ is the co-moving distance to the centre of the survey volume
which has a co-moving depth $\Delta D$. Here $n(k_{\perp})$ is the
density of baselines which observe a wave vector with transverse
component $k_{\perp}$, and $\theta$ is the angle between the mode
$\vec{k}$ and the line of sight. The thermal noise component (first
term) is proportional to the sky temperature, where $T_{\rm sky} \sim
250 \left( \frac{1+z}{7}\right)^{2.6}$\,K at the frequencies of
interest. The second term corresponds to sample variance. The overall
sensitivity is
\begin{equation}
\Delta P_{21}(k,\theta) = \delta P_{21}(k,\theta)/\sqrt{N_c(k,\theta)},
\end{equation}
where $N_c(k,\theta)$ denotes the number of modes observed in a
$k$-space volume $d^3k$ (only modes whose line-of-sight components fit
within the observed band-pass are included). In terms of the
$k$-vector components $k$ and $\theta$, $N_c = 2\pi k^2 \sin\theta dk
d\theta \mathcal{V}/(2 \pi)^3$ where $\mathcal{V} = D^2 \Delta D
(\lambda^2/A_e)$ is the observed volume. Taking the spherical average
over bins of $\theta$, the sensitivity to the 21-cm PS is
\begin{equation}
\frac{1}{[\Delta P_{21}(k)]^2}=\sum_{\theta}\frac{1}{[\Delta P_{21}(k,\theta)]^2}.
\end{equation}

The spherically averaged sensitivity curves for the MWA (within bins
of $\Delta k=k/10$) are plotted as the dotted lines in the lower
panels of Figure~\ref{fig5}. The sensitivity as a ratio of the PS
computed without a galactic H\,{\sevensize I} contribution ($\Delta
P_{21}/P_{21}$) is plotted in the upper panels of Figure~\ref{fig5}
(again as dotted lines). These estimates illustrate that late in
reionisation, the effect of the galactic H\,{\sevensize I} will be at
a level above the sensitivity of the first generation low-frequency
arrays. However, the low resolution of the MWA would mean that the
shape-change of the PS owing to the galactic H\,{\sevensize I} would
not be detected until very close to overlap. The corresponding
sensitivity curves for the MWA5000 are plotted as the thin grey lines
in Figure~\ref{fig5}. The larger collecting area of the MWA5000 would
allow the effect of galactic H\,{\sevensize I} to be detected out to
higher redshifts and at smaller scales.

\subsection{Sensitivity to the 21cm-galaxy cross PS}

To compute the sensitivity of the MWA to the 21cm-galaxy cross PS ($P_{\rm 21,g}$) we follow the discussion of Furlanetto \& Lidz~(2007). The sensitivity to a particular mode is 
\begin{equation}
2[\delta P_{\rm 21,g}(k,\theta)]^2 = [P_{\rm 21,g}(k,\theta)]^2 + \delta P_{21}(k,\theta)\delta P_{\rm g}(k,\theta),
\end{equation}
where $\delta P_{\rm g}(k,\theta)=b^2P_{\delta}(k,\theta)+n_{\rm
  g}^{-1}$ is the uncertainty in the galaxy PS, and $P_{\delta}$ is
the underlying mass PS\footnote{Note that for consistency with our
  estimate of the 21-cm PS, we are neglecting redshift space
  distortions in calculation of the galaxy PS.} at redshift $z$. In
the second term $n_{\rm g}$ is the density of galaxies (which we
approximate as $n_{\rm g}=M_{\rm g}dn/dM_{\rm g}$, where $dn/dM$ is
the Press-Schechter~1976) mass function and $M_{\rm g}$ is the halo
mass of galaxies in the survey\footnote{Note that we have neglected
  redshift errors when computing the Poisson component of the
  uncertainty in the galaxy PS.}. After calculation of the total sensitivity 
\begin{equation}
\Delta P_{\rm 21,g}(k,\theta) = \delta P_{\rm 21,g}(k,\theta)/\sqrt{N_c(k,\theta)},
\end{equation}
and taking the spherical average over bins of $\theta$ as before, the
sensitivity to the 21cm-galaxy cross PS is
\begin{equation}
\frac{1}{[\Delta P_{21,g}(k)]^2}=\sum_{\theta}\frac{1}{[\Delta P_{21,g}(k,\theta)]^2}.
\end{equation}

The spherically averaged sensitivity curves for the MWA (within bins
of $\Delta k=k/10$) are plotted as the dotted lines in the central
panels of Figure~\ref{fig6}. The sensitivity as a ratio of the
fiducial IGM only PS ($\Delta P_{\rm 21,g}/P_{\rm 21,g}$) is also
plotted in the upper panels of Figure~\ref{fig6}. Here we have assumed
that galaxies have been detected down to a resolution limit
corresponding to a host mass of $M_{\rm g}=10^{11}M_\odot$ over the
full MWA field. These sensitivity estimates illustrate that the effect
of the galactic H\,{\sevensize I} on the cross-correlation between
galaxies and 21-cm emission could be detected with the first
generation low-frequency arrays. Moreover the resolution of the MWA
would be sufficient to detect the sign-change of the 21cm-galaxy cross
PS at the scale of the H\,{\sevensize II} regions late in
reionisation. The corresponding sensitivity curves for the MWA5000 are
plotted as the thin grey lines in Figure~\ref{fig6}. The larger
collecting area of the MWA5000 would greatly increase the significance
with which the effect of galactic H\,{\sevensize I} could be measured.

\section{conclusion}
\label{conclusion}

In this paper we have investigated the impact of H\,{\sevensize I} in
galaxies on the statistics of 21-cm fluctuations using analytic and
semi-numerical models.  Our models are unable to self-consistently
compute the galactic H\,{\sevensize I} content of galaxies prior to
the end of reionisation. As an input to our model we have therefore
assumed that during the reionisation era 2\% of hydrogen is in the
form of H\,{\sevensize I} and located within galaxies. This number is
motivated by observations of the mass weighted fraction of cosmic
hydrogen in H\,{\sevensize I} after the end of reionisation, which is
dominated by damped absorption systems and which has a constant value
of $\sim2\%$ between $z\sim1$ and $z\sim5$. Our modelling shows that
this assumption results in a reduction of 10-20\% in the amplitude of
21-cm fluctuations over a range of spatial scales, and over a large
fraction of the reionisation era. In addition to the amplitude of
21-cm fluctuations we have also modelled the cross-correlation between
galaxies and 21-cm emission. We find that the inclusion of galaxies
decreases the amplitude of the cross-correlation by a few tens of
percent. In addition, the cross-correlation between galaxies and 21-cm
emission will change sign at the end of the reionisation era,
providing an alternative avenue to pinpoint the end of reionisation.

We find that our analytic estimates become less applicable once
H\,{\sevensize II} regions become a dominant feature of the IGM. We
have therefore supplemented our analytic estimates with semi-numerical
modelling of the three dimensional ionisation structure of the IGM. We
find that the qualitative conclusions from our analytic calculations
are vindicated by this modelling. In addition to the above estimates
of auto and cross-correlation, we have used our semi-numerical model
to compute 21-cm power spectra, and 21cm-galaxy cross power spectra.
The effect of galactic H\,{\sevensize I} is to change the shape of the
21-cm PS rather than just the amplitude. In particular, because the
galactic H\,{\sevensize I} is biased towards H\,{\sevensize II}
regions, we find that the H\,{\sevensize II} region induced {\em
  shoulder} in the PS is less prominent when the contribution of
galactic H\,{\sevensize I} is considered.

The amplitude of the 21-cm PS is modified in a scale dependent way by
up to 20\% when a 2\% galactic H\,{\sevensize I} fraction is included.
Of course the galactic H\,{\sevensize I} fraction is very uncertain at
high redshift. However we find that the fractional modification of the
21-cm fluctuation statistics is approximately proportional to the
density of galactic H\,{\sevensize I}. Therefore, if we assume a value
lower(higher) than the 2\% observed at $z<5$, then our results for the
error introduced through neglect of galactic H\,{\sevensize I} will be
over(under)estimated. On the other hand, the change of sign in the
cross-correlation between galaxies and the 21-cm signal is robust to
our assumptions.

We find that the inclusion of galactic H\,{\sevensize I} lessens
the amplitude of the anti-correlation between galaxies and 21-cm
emission since a fraction of H\,{\sevensize I} is now
co-located with the galaxies inside the H\,{\sevensize II} regions. In
addition, the cross PS changes sign on small scales, which reflects the
correlation of galaxies with the galactic H\,{\sevensize I} inside the
H\,{\sevensize II} regions, where no power is contributed in 21-cm
fluctuations of the IGM. Thus, the scale at which the 21cm-galaxy
cross PS changes sign could be used to probe the scale of
H\,{\sevensize II} regions late in the reionisation era.

We have estimated the sensitivity of the MWA to the spherically
averaged 21-cm PS and 21cm-galaxy cross PS. Our calculations
illustrate that the effect of the galactic H\,{\sevensize I} on 21-cm
fluctuations is at a level that would be significant with respect to
the sensitivity of the MWA late in reionisation. However the low
resolution of the MWA would mean that the shape change of the PS owing
to the galactic H\,{\sevensize I} could not be detected until very
close to overlap. On the other hand, when combined with a suitable
galaxy redshift survey, the resolution of the MWA would be sufficient
to detect the sign change of the 21cm-galaxy cross PS at the scale of
the H\,{\sevensize II} regions late in reionisation. A followup
telescope comprising 10 times the collecting area of the MWA would
measure the effects of galactic H\,{\sevensize I} with high
significance.

In summary, our modelling shows that the H\,{\sevensize I} content of
the galaxies that reionise the universe provides a significant
contribution to the statistics of 21-cm fluctuations. The galactic
H\,{\sevensize I} contribution to the 21-cm intensity will therefore
need to be considered in detailed modelling of the 21-cm intensity PS
in order to correctly interpret measurements from the next generation
of low-frequency arrays.

{\bf Acknowledgements} The research was supported by the Australian
Research Council (JSBW). LW and PMG acknowledge the support of
Australian Postgraduate Awards. SPO acknowledges support from NASA
grant NNG06GH95G.

\newcommand{\noopsort}[1]{}

\label{lastpage}

\begin{thebibliography}{}

\bibitem[Barkana \& Loeb(2005)]{2005ApJ...624L..65B} Barkana, R., \& Loeb, 
A.\ 2005, \apjl, 624, L65 

\bibitem[Barkana(2007)]{2007MNRAS.376.1784B} Barkana, R.\ 2007, \mnras, 
376, 1784 

\bibitem[Bharadwaj \& Ali(2005)]{2005MNRAS.356.1519B} Bharadwaj, S., \& 
Ali, S.~S.\ 2005, \mnras, 356, 1519 

\bibitem[Bond et al.(1991)]{1991ApJ...379..440B} Bond, J.~R., Cole, S., 
Efstathiou, G., \& Kaiser, N.\ 1991, \apj, 379, 440 

\bibitem[]{} 
Bowman, J., Morales, M., Hewitt, J., 2005, \apj, 638, 20

\bibitem[Bowman et al.(2007)]{2007ApJ...661....1B} Bowman, J.~D., Morales, 
M.~F., \& Hewitt, J.~N.\ 2007, \apj, 661, 1 

\bibitem[]{} 
Dijkstra, M., Haiman, Z., Rees, M.~J., \& Weinberg, D.~H., { Astrophys. J.}, { 601}, 666-675 (2004)

\bibitem[]{} 
Efstathiou, G., { Mon. Not. R. Astron. Soc.}, { 256}, 43-47 (1992)

\bibitem[Fan et al.(2006)]{2006AJ....132..117F} 
Fan, X., et al.\ 2006, \aj, 132, 117 

\bibitem[Furlanetto et al.(2004)]{2004ApJ...613...16F} 
Furlanetto, S.~R., Zaldarriaga, M., \& Hernquist, L.\ 2004, \apj, 613, 16 

\bibitem[Furlanetto et al.(2006)]{2006PhR...433..181F} 
Furlanetto, S.~R., Oh, S.~P., \& Briggs, F.~H.\ 2006, Physics Reports, 433, 181 

\bibitem[Furlanetto 
\& Lidz(2007)]{2007ApJ...660.1030F} Furlanetto, S.~R., \& Lidz, A.\ 2007, \apj, 660, 1030 

\bibitem[Geil \& Wyithe(2008)]{2008MNRAS.386.1683G} 
Geil, P.~M., \& Wyithe, J.~S.~B.\ 2008, \mnras, 386, 1683 

\bibitem[Gnedin \& Fan(2006)]{2006ApJ...648....1G} 
Gnedin, N.~Y., \& Fan, X.\ 2006, \apj, 648, 1 

\bibitem[Iliev et al.(2008)]{2008MNRAS.384..863I} 
Iliev, I.~T., Mellema, G., Pen, U.-L., Bond, J.~R., \& Shapiro, P.~R.\ 2008, \mnras, 384, 863 

\bibitem[]{}
Kashikawa, N., et al. 2006, \apj, 648, 7 

\bibitem[Komatsu et al.(2008)]{2008arXiv0803.0547K} Komatsu, E., et al.\ 
2008, ArXiv e-prints, 803, arXiv:0803.0547 

\bibitem[Kramer et al.(2006)]{2006ApJ...649..570K} Kramer, R.~H., Haiman, 
Z., \& Oh, S.~P.\ 2006, \apj, 649, 570 

\bibitem[Lidz et al.(2008)]{2008arXiv0806.1055L} Lidz, A., Zahn, O., 
Furlanetto, S., McQuinn, M., Hernquist, L., 
\& Zaldarriaga, M.\ 2008, ArXiv e-prints, 806, arXiv:0806.1055 

\bibitem[McQuinn et al.(2006)]{2006ApJ...653..815M} 
McQuinn, M., Zahn, O., 
Zaldarriaga, M., Hernquist, L., \& Furlanetto, S.~R.\ 2006, \apj, 653, 815 

\bibitem[Mesinger \& Furlanetto(2007)]{2007ApJ...669..663M} 
Mesinger, A., \& Furlanetto, S.\ 2007, \apj, 669, 663 

\bibitem[Mo \& White(1996)]{1996MNRAS.282..347M} 
Mo, H.~J., \& White, S.~D.~M.\ 1996, \mnras, 282, 347 

\bibitem[Morales et al.(2006)]{2006ApJ...648..767M} Morales, M.~F., Bowman, 
J.~D., \& Hewitt, J.~N.\ 2006, \apj, 648, 767 

\bibitem[]{}
Press, W., Schechter, P., 1974, {ApJ.}, {187}, 425

\bibitem[Prochaska et al.(2005)]{2005ApJ...635..123P} 
Prochaska, J.~X., Herbert-Fort, S., \& Wolfe, A.~M.\ 2005, \apj, 635, 123 


\bibitem[]{} 
Quinn, T., Katz, N., \& Efstathiou, G., {278}, L49-L54 (1996)

\bibitem[Shapiro et al.(2004)]{2004MNRAS.348..753S} Shapiro, P.~R., Iliev, 
I.~T., \& Raga, A.~C.\ 2004, \mnras, 348, 753 

\bibitem[]{} 
Thoul, A.~A., \& Weinberg, D.~H., { Astrophys. J.}, {465}, 608-116 (1996)

\bibitem[]{}
White, R., Becker, R., Fan, X., Strauss, M., 2003, {Astron J.}, {126}, 1 

\bibitem[Wyithe \& Loeb(2007)]{2007MNRAS.375.1034W} 
Wyithe, J.~S.~B., \& Loeb, A.\ 2007, \mnras, 375, 1034 

\bibitem[Wyithe et al.(2008)]{2008MNRAS.383.1195W} Wyithe, J.~S.~B., Loeb, 
A., \& Geil, P.~M.\ 2008, \mnras, 383, 1195 

\bibitem[Wyithe 
\& Morales(2007)]{2007MNRAS.379.1647W} Wyithe, J.~S.~B., \& Morales, M.~F.\ 2007, \mnras, 379, 1647 

\bibitem[Zahn et al.(2007)]{2007ApJ...654...12Z} 
Zahn, O., Lidz, A., McQuinn, M., Dutta, S., Hernquist, L., Zaldarriaga, M., 
\& Furlanetto, S.~R.\ 2007, \apj, 654, 12 

\bibitem[Zaldarriaga et al.(2004)]{2004ApJ...608..622Z} 
Zaldarriaga, M., Furlanetto, S.~R., \& Hernquist, L.\ 2004, \apj, 608, 622 


\end{thebibliography}
\end{document}